\documentclass[12pt] {article}
\usepackage{graphicx}
\usepackage{epsfig}
\begin {document}
\parindent=15pt
\vskip .8 truecm
\begin{center}
{\bf NUCLEAR RADII CALCULATIONS IN VARIOUS THEORETICAL APPROACHES
FOR NUCLEUS-NUCLEUS INTERACTIONS}\\ 
\vspace{.5cm}
C. Merino$^a$, I. S. Novikov$^b$, and Yu. M. Shabelski$^c$ \\
\vspace{.5cm}
$^a$ Departamento de F\'\i sica de Part\'\i culas, Facultade de F\'\i sica, \\
and Instituto Galego de F\'\i sica de Altas Enerx\'\i as (IFGAE), \\
Universidade de Santiago de Compostela, \\
15782 Santiago de Compostela, Galiza, Spain \\
merino@fpaxp1.usc.es \\
\vspace{.1cm}
$^b$ Department of Physics and Astronomy, \\
Western Kentucky University, \\
1906 College Heights Blvd, \#11077, Bowling Green, \\
KY 42101-1077, USA\\
ivan.novikov@wku.edu \\
\vspace{.1cm}
$^c$ Petersburg Nuclear Physics Institute, \\
Gatchina, St.Petersburg 188350, Russia \\
shabelsk@thd.pnpi.spb.ru \\
\end{center}
\vspace{.1cm}

\begin{abstract}

The information about sizes and nuclear density distributions in unstable (radioactive)  nuclei is usually extracted from the data on interaction of radioactive nuclear beams  with a nuclear target. We show that in the case of nucleus-nucleus collisions the values of the parameters depend rather strongly on the considered theoretical approach and on the assumption about the parametrization of the nuclear density distribution. The obtained values of root-mean-square radii ($\rm R_{rms}$) for stable nuclei with atomic weights A = 12$-$40 vary by approximately 0.1 fm when calculated in the optical approximation, in the rigid target approximation, and using the exact expression of Glauber Theory. We present several examples of $\rm R_{rms}$ radii calculations using these three theoretical approaches and compare these results with the data obtained from electron-nucleus scattering.

\end{abstract}

\vskip .3cm
PACS numbers: 21.10.Gv, 25.60.Dz, 25.70.Bc
\vskip .3cm

\pagebreak

\section{Introduction}

The root-mean-square nuclear matter radii ($\rm R_{rms}$) and the density distributions contain an important insight on nuclear potentials and nuclear wave functions. 

In the case of stable nuclei the most precise information comes from the data on elastic scattering of fast particles on nuclear targets,~ref.~\cite{Elt}. The comparison of the data on electron and proton elastic scattering gives separate information about proton and neutron distributions in the nucleus,~refs.~\cite{CLS,ABV}. The Glauber Theory,~refs.~\cite{Gl1,Sit1,Gl2}, is typically used to analyze experiments on interactions with nuclei at energies higher than several hundred MeV. The obtained values of parameters for nuclear matter density and for charge density distributions are summarized in~refs.~\cite{CLS,ABV,DeDe}.

Currently, the only way to study density distribution of the unstable (radioactive) nuclei is to scatter them off a target. There are a number of technical difficulties in the use of hydrogen as target for these experiments, so targets are more often made out of different stable nuclei (for example, $^{12}$C,~refs.~\cite{Oza,Oza1}). In this case the analytical calculation of all Glauber diagrams for nucleus-nucleus interactions becomes impossible and some approximative approaches have to be used. The results for nuclear radii and, possibly, for nuclear density distributions depend on the considered theoretical approach. These problems were already considered for $^6$He and $^{11}$Li scattering on $^{12}$C target in refs.~\cite{Gar,AlLo}. 

In the present paper we extract the values of nuclear $\rm R_{rms}$ from the data refs.~\cite{Oza,Oza1} on interaction cross sections of stable nuclei with atomic weights A = 12$-$40 colliding with a carbon target. First, we show that the obtained values of $\rm R_{rms}$ depend on the parameterization of the nuclear matter distribution that is used. Then, by assuming that the nuclear density in this range of atomic weights can be described by the Woods-Saxon expression, we extract the values of $\rm R_{rms}$ radii from the same experimental data in the frame of three different theoretical approaches: the optical approximation, the rigid target approximation, and the exact expression of the Glauber Theory. Calculations in the  Glauber Theory framework are performed using a Monte Carlo simulation technique. Finally, we show that the accounting for the non-zero range of NN interaction significantly changes the $\rm R_{rms}$ values.

Our goal is to demonstrate the numerical level of uncertainties which can appear in the analysis of the interaction of unstable nuclei.  In general, the optical approximation and the rigid target approximation result in smaller values of $\rm R_{rms}$ when compared with the values extracted from the electron scattering data,~ref.~\cite{DeDe}. The values obtained in the framework of the exact Glauber Theory expression are in better agreement with the electron scattering data. 

\section{Elastic nucleus-nucleus scattering in the Glauber theory}

In the Glauber theory the elastic scattering amplitude of nucleus $A$ on nucleus $B$ with momentum transfer $q$ can be written in the frame where $B$-nucleus is a fixed target as in refs.~\cite{PaRa,BrSh}:
\begin{equation}
F^{el}_{AB}(q) = \frac{ik}{2\pi} \int d^2b\hspace{0.1cm} e^{iqb}\cdot [1 - S_{AB}(b)] \;,
\end{equation}
where $k$ is the incident momentum of one nucleon in $A$-nucleus in laboratory frame, $b$ is the impact parameter, and
\begin{equation}
S_{AB}(b) = \langle A \vert \langle B \vert \left\{
\prod_{\stackrel{i \in A}{j \in B}}  [1 - \Gamma_{NN}(b + u_i - s_j)] \right\} \;
\vert B \rangle \vert A \rangle \;,
\end{equation}
with
\begin{equation}
\Gamma_{NN}(b + u_i - s_j) = \frac{1}{2 \pi ik} \int d^2q
\hspace{0.1cm} e^{-iq(b + u_i - s_j)}\cdot f_{NN}(q) \;.
\end{equation}
Here $u_i$ and $s_j$ are the transverse coordinates of nucleons, and $f_{NN} (q)$ is the amplitude of elastic nucleon-nucleon scattering, that can be parameterized as
\begin{equation}
f_{NN} (q) = \frac{ik\sigma}{4 \pi} \exp \left( -\frac12 \beta q^2\right).
\end{equation}
Here $\sigma$ is the total NN cross section, $\beta$ is the slope parameter of  NN elastic scattering. We neglect the real part of $f_{NN} (q)$ since it gives negligible ($\sim 1\%$) contribution to the reaction cross section. In the following calculations we use the values $\sigma$ = 43 mb and $\beta$ = 6 GeV$^{-2}$.

Contrary to the case of hadron-nucleus interaction, we can not integrate the Eq.~(2) analytically, even with the standard assumption that nuclear densities $\rho(r_1,...,r_A)$ in both $A$ and $B$ nuclei are the normalized products of one-nucleon densities $\rho(r_i)$:
\begin{equation}
\rho(r_1,...,r_A) = \prod^A_{i=1} \rho_A(r_i) \;, \;\;
\rho(r_1,...,r_B) = \prod^B_{i=1} \rho_B(r_i) \;, \;\;
\; \int d^3r_i \rho(r_i) =1 \;.
\end{equation}

To make the problem manageable, one can retain only part of all  contributions in the expansion of the product in Eq.~(2), that corresponding to the contributions characterized by large combinatorial factors. The leading graphs correspond to the so-called optical approximation, ref.~\cite{CzMa}, in which one sums up the contributions with no more than one scattering for each nucleon. In other words, only those products of amplitudes $\Gamma_{NN}(b + u_i - s_j)$  in Eq.~(2) are taken into accounts which have different indices $i$, $j$. This approximation corresponds to the summation of diagrams shown in Figs.~1a, 1b, 1c,...  In this approximation diagram which describes $n$-fold interaction has a combinatorial factor $_{A}C_{n} \; \cdot \; _{B}C_{n}$. To avoid crowding of lines in Fig.~1 we have only shown the nucleon participants from the nucleus $A$ (upper dots) and $B$ (lower dots) with the links standing for interacting amplitudes, and we have not plotted the nucleon-spectators.

\begin{figure}[htb]
\begin{center}
\mbox{\psfig{file=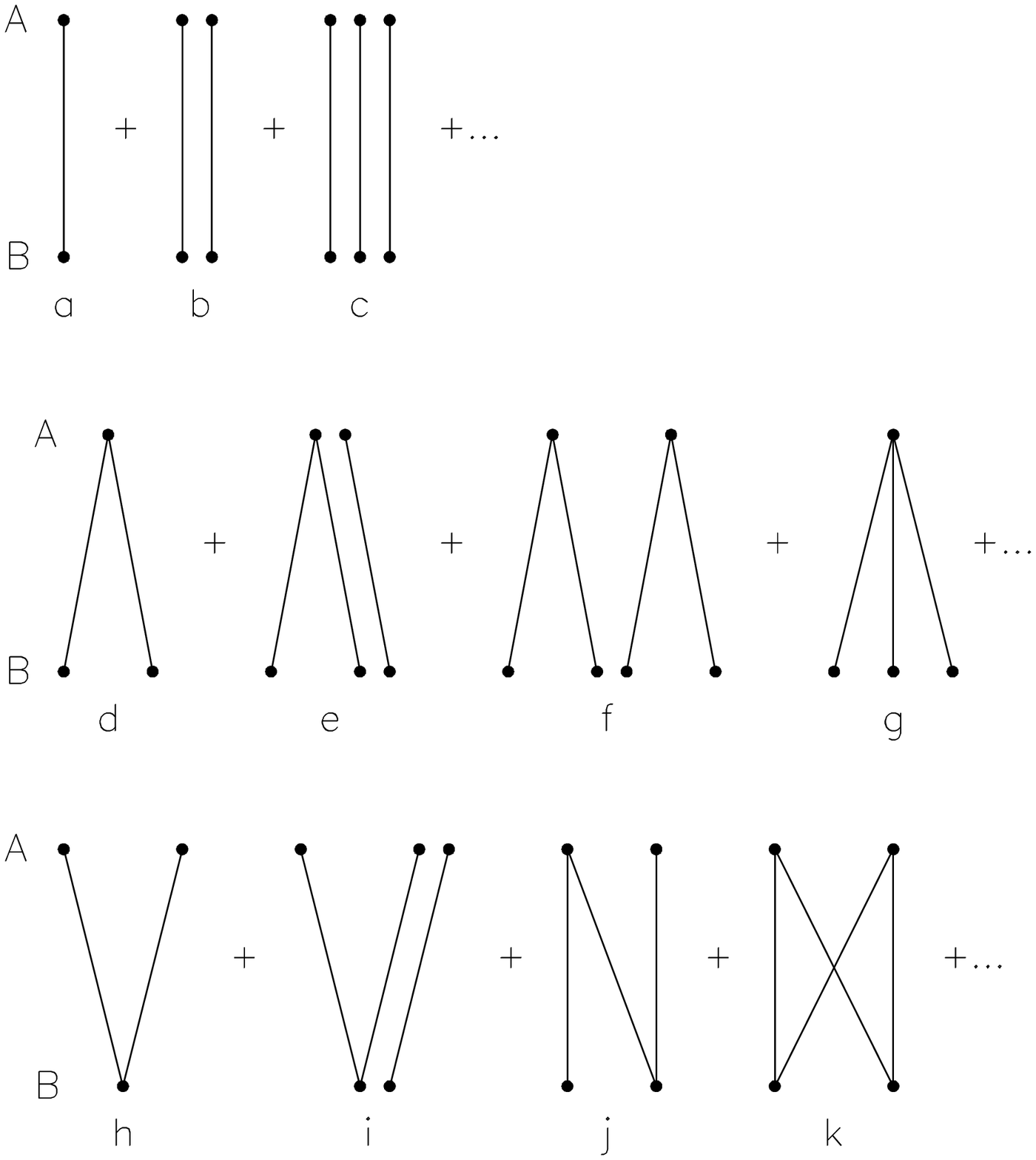,width=0.7\textwidth}} \\
Fig.~1. Diagrams of the interaction of two nuclei, $A$ and $B$, taken
into account \\
\hspace{-3.cm }in different approximations of the multiple scattering theory.
\end{center}
\end{figure}

In the optical approximation averaging $\langle A \vert ... \vert A \rangle$ and $\langle B \vert ... \vert B \rangle$ of the product  $\prod_{\stackrel{i \in A}{j \in B}}  [1 - \Gamma_{NN}(b + u_i - s_j)]$ can be substituted with the averaging of the $[1 - \Gamma_{NN}(b + u_i - s_j)]$:
\begin{equation}
S_{AB}^{opt}(b) = \prod_{i,j} \langle A \vert \langle B \vert
[1 - \Gamma_{NN}(b + u_i - s_j)] \; \vert B \rangle \vert A \rangle \;.
\end{equation}
Using the standard assumptions of the multiple scattering theory one obtains:
\begin{equation}
S_{AB}^{opt}(b) = \left [1 - \frac1A T_{opt}(b) \right ]^A \approx
\exp [- T_{opt}(b)] \;,
\end{equation}
where
\begin{equation}
T_{opt}(b) = \frac{\sigma}{4 \pi \beta} \int d^2b_1 d^2b_2 T_A(b_1)\cdot T_B(b_2)\cdot
\exp\left[-\frac{(b+b_1-b_2)^2}{2\beta}\right] \;,
\end{equation}
with 
\begin{equation}
T_A(b) = A \int^{\infty}_{-\infty} dz \rho_A\left(\sqrt{b^2+z^2} \right) \;.
\end{equation}

Neglecting the NN interaction range in comparison to the nuclear radii, we have
\begin{equation}
T_{opt}(b) = \frac{\sigma}{2}\cdot \int d^2b_1 T_A(b-b_1)\cdot T_B(b_1) \;.
\end{equation}

In the diagramatic language Eq.~(2) accounts for all possible intermediate states of nucleons between the interactions, as it is shown in Fig.~2a, while the optical approximation, Eq.~(6), would correspond to the interactions with only one pole (nuclear ground state) in the both $A$ and $B$ intermediate states, ref.~\cite{BrSh} (see Fig.~2c).
\begin{figure}[htb]
\begin{center}
\vskip -7cm
\mbox{\psfig{file=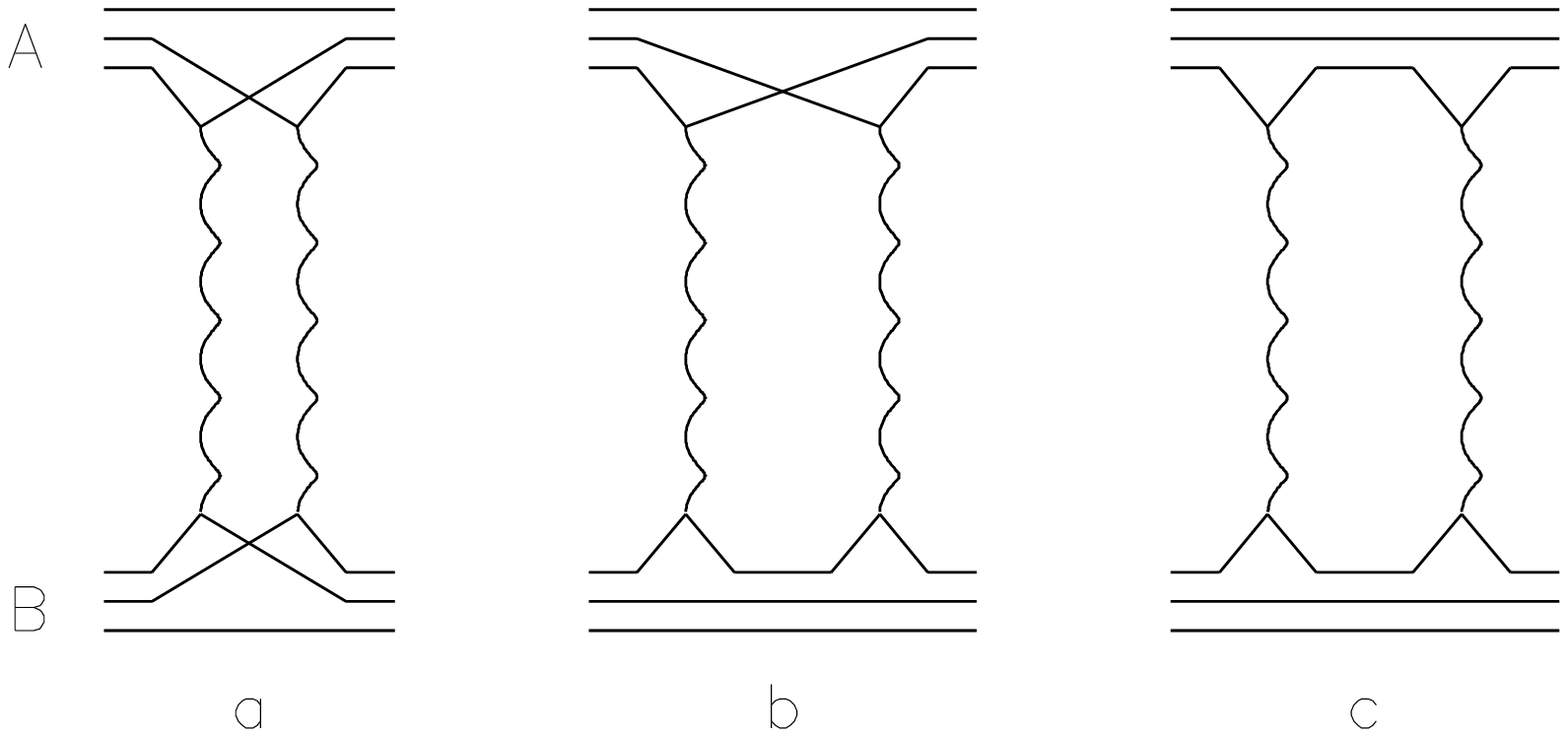,width=0.8\textwidth}} \\
Fig.~2. Two-fold interaction of two nuclei in the multiple scattering
theory (a), in the\\
\hspace{3.cm}rigid target approximation (b) and in the optical approximation (c).
\end{center}
\end{figure}

Unfortunately, numerical calculations in ref.~\cite{FV} (see also ref.~\cite{Sh2} for the case of collisions of very light nuclei) demonstrate that the optical approximation is not accurate enough even for the integrated cross sections. The difference with the data amounts for $\sim 10-15$ \% in $\sigma_{AB}^{tot}$ and it is even greater for differential cross sections, ref.~\cite{FV}. This disagreement can be explained by the fact that series with smaller combinatorial factors in Eq.~(2) give significant global corrections to the optical approximation results. As a matter of fact, the terms of the series are alternating in sign, so, due to the cancelations of terms with opposite signs, the final sums of these series can take very different values. Thus, some classes of diagrams with non-leading combinatorial factors give significant contributions to the final total value. 

The rigid target (or rigid projectile) approximation, refs.~\cite{Alk,VT}, is more explicit than the optical approximation. It corresponds to averaging $\langle B \vert ... \vert B \rangle$ inside the product in Eq.~(2):
\begin{equation}
S_{AB}^{r.g.}(b) = \langle A \vert \left \{ \prod_{i,j} \vert \langle B \vert [1 - \Gamma_{NN}(b + u_i - s_j)] \; \vert B \rangle \right \} \vert A \rangle = [T_{r.g.}(b)]^A \;,
\end{equation}
where
\begin{equation}
T_{r.g.}(b) = \frac1A \int d^2b_1 T_A(b_1) \exp \left \{-\frac{\sigma}{4 \pi \beta} \int d^2b_2 T_B(b_2) \exp\left[-\frac{(b+b_1-b_2)^2}{2\beta}\right]   \right \} \;.
\end{equation}
Neglecting the NN interaction range in comparison to the nuclear radii we have:
\begin{equation}
T_{r.g.}(b) = \frac1A \int d^2b_1 T_A(b_1 - b)\cdot \exp \left [ - \frac{\sigma}{2} T_B(b_1) \right ] \;.
\end{equation}

This approximation corresponds to the sum of the diagrams in  Figs.~1a, 1b, 1c, ... and the correction diagrams in Figs.~1d, 1e, 1f, 1g,... Diagrams in  Figs.~1d, 1e, 1f, 1g,... represent the case when each nucleon from the nucleus $A$ can interact several times, but all interacting nucleons from $B$ are still different.  Each correction diagram which describes $n$-fold interaction has a combinatorial factor smaller than the combinatorial factor leading diagrams $_{A}C_{n} \; _{B}C_{n}$ (Fig. 1a, 1b, 1c, ...). Although, due to the obvious asymmetry in contributions of the two nuclei such approach can be theoretically justified in the limit $A/B \ll 1$, i.e. for C-Pb, or S-U collisions, this approximation can be used sometimes in the case of heavy ion collisions with equal atomic weights.

Further corrections to the elastic amplitude (some of them are shown in Figs.~1h, 1i, 1j, 1k,...) have been considered in refs.~\cite{Andr1,Andr2,MBra,BoKa}. However, the results of such corrections are rather complicated for practical use.

The possibility to obtain the Glauber Theory results without any simplification by the direct calculation of Eq. (2) using Monte Carlo simulation was first suggested in refs.~\cite{ZUS,Shm}. This method was used for numerical calculations in refs.~\cite{Gar,AlLo,Sh11}. The simplest algorithm considering the values of coordinates uniformly distributed in the interaction region can not be applied here because in most cases several coordinates have values corresponding to very small nuclear density. 

The algorithm proposed by Metropolis {\it et al} in~\cite{Metro} allows to generate a set of coordinates which re distributed according to a pre-defined distribution. The Metropolis method is as follows:
\begin{enumerate}
\item The initial coordinate $s_i$ is randomly generated from the appropriate interval
\item To obtain next coordinate, a shift $\Delta s_i$ is randomly generated and then added to the initial coordinate
\item New coordinate is accepted when the ratio $r = \rho(s_i + \Delta s_i)/\rho(s_i) >1$
\item If the ratio $r<1$, new coordinate is accepted only when $r>x$, where $x$ is a new random number $x$ from $[0,1]$ interval. Otherwise new coordinate is not accepted
\end{enumerate}

Generated set of coordinates was used to calculate average value of $\prod_{ij} [1 - \Gamma_{NN}(b + u_i - s_j)]$. Finally, we calculated $S_{AB}(b)$ with several different sets of nucleon coordinates.

\section{Monte Carlo simulation of reaction nucleus-nucleus cross sections}

The total inelastic (reaction) cross section for the collisions of nuclei A and B, $\sigma^{(r)}_{AB}$, is equal to the difference of the total interaction cross section $\sigma^{tot}_{AB}$ and the integrated elastic scattering cross section, $\sigma^{el}_{AB}$:
\begin{equation}
\sigma^{(r)}_{AB} = \sigma^{tot}_{AB} - \sigma^{el}_{AB} = \int d^2b [1 - \vert S_{AB}(b)\vert^2] \;.
\end{equation}
where $\sigma^{tot}_{AB}$ and $\sigma^{el}_{AB}$ are
\begin{equation}
\sigma^{tot}_{AB} = \frac{4 \pi}k \mathrm{Im} \, F^{el}_{AB}(q=0) = 2 \int d^2b [1 - S_{AB}(b)] \; ,
\end{equation}
\begin{equation}
\sigma^{el}_{AB} =  \int d^2b [1 - S_{AB}(b)]^2 \;,
\end{equation}
Only interaction cross section $\sigma^{(I)}_{AB}$ have been measured in refs.~\cite{Oza,Oza1}. The difference between $\sigma^{(r)}_{AB}$ and  $\sigma^{(I)}_{AB}$ is that the reaction cross sections include the cross sections of all processes except of the elastic scattering AB $\to$ AB, whereas the interaction cross sections do not include the processes with a target nucleus exitation or with disintegration AB $\to$ AB$^*$ ($B^* \neq B$), hence, $\sigma^{(I)}_{AB} < \sigma^{(r)}_{AB}$.  However, the difference between $\sigma^{(I)}_{AB}$ and $\sigma^{(r)}_{AB}$ has been estimated to be less than a few percent for beam energies higher than several hundred MeV per nucleon, refs.~\cite{KSh,Oga}. That allows us to neglect in the present paper the difference between $\sigma^{(I)}_{AB}$ and $\sigma^{(r)}_{AB}$ and compare our calculations of $\sigma^{(r)}_{AB}$ to the experimental data from refs.~\cite{Oza,Oza1} on $\sigma^{(I)}_{AB}$ in the same way as it was done in refs.~\cite{Oza,AlLo}.

For the numerical calculations it is necessary to use an expression for the nuclear matter density distributions in the colliding nuclei. The most detailed information about these distributions comes from the data on the differential elastic scattering cross sections on nuclear targets. The nucleon density in the light (A$ \leq 20$) nuclei can be described by a harmonic oscillator (HO) density distribution, ref.~\cite{Oza}:
\begin{equation}
\rho_A (r) = \rho_1\cdot \left(1+\frac{A/2-2}{3}\cdot \left( \frac{r}{\lambda}\right)^2 \right)\cdot
\exp\left(-\frac{r^2}{\lambda^2}\right) ,
\end{equation}
where $\lambda$ is the nucleus size parameter and $\rho_{1}$ is the normalization 
constant, while the nuclear density distributions in not very light nuclei can be reasonably 
described by the Woods-Saxon expression
\begin{equation}
\rho_A (r) = \frac{\rho_1}{1+\exp\left((r-c)/a\right)} \;.
\end{equation}
Here $\rho_{1}$ is the normalization constant, $c$ is a parameter measuring the nuclear size, and $a$ is related to the diffuseness of the surface, in other words, to the thickness of the nuclear skin. The parameter $c$ shows the value of $r$ at which $\rho(r)$ decreases by a factor 2 compared to $\rho(r=0)$, $\rho(r=c) = \frac{1}{2}\rho (r = 0)$. The value of $a$ determines the distance $r$ ($r = 4a \ln 3 \sim 4.4 a$) at which $\rho(r)$ decreases from $0.9\rho(r = 0)$ to  $0.1\rho(r = 0)$.

From the experimental data on $\sigma^{(I)}_{AB}$ it is possible to determine only one parameter of nuclear matter distribution, let's say R$_{\rm rms}$:
\begin{equation}
{\rm R_{rms}} = \sqrt{\langle r^2_A \rangle } \;, \;\;
\langle r^2_A \rangle = \int r^2 \rho_A (r) d^3r \;.
\end{equation}
It is needed to note that the value of ${\rm R_{rms}}$ is rather smaller than the standard nuclear radius $R_A \simeq 1.2 A^{1/3}$, fm. For example, in the case of uniform nuclear density with radius $R_A$, $R_A = \sqrt{\frac{5}{3}\langle r^2_A \rangle}$.

The reaction cross section of  $^{12}$C-$^{12}$C interaction as a function of the ${\rm R_{rms}}$ of the $^{12}$C nucleus was calculated in different theoretical approximations for various nucleon density distributions. The results of the calculations are shown in Figure 3.  In these calculations, which correspond to energies of 800-1000 MeV per projectile nucleon, the total NN cross section value, averaged over $pp$ and $pn$ interactions, $\sigma^{tot}_{NN} =$ 43 mb was used.  The nuclear matter distribution parameters $\lambda$, for harmonic oscillator (HO) density in Eq.~(17), and $c$, for the Woods-Saxon density in Eq.~(18), have been fitted to obtain the required ${\rm R_{rms}}$, whereas the parameter $a$ in Eq.~(15) has been fixed to the value $a = 0.54$ fm. 

\begin{figure}[htb]
\begin{center}
\vskip-0.5cm
\mbox{\psfig{file=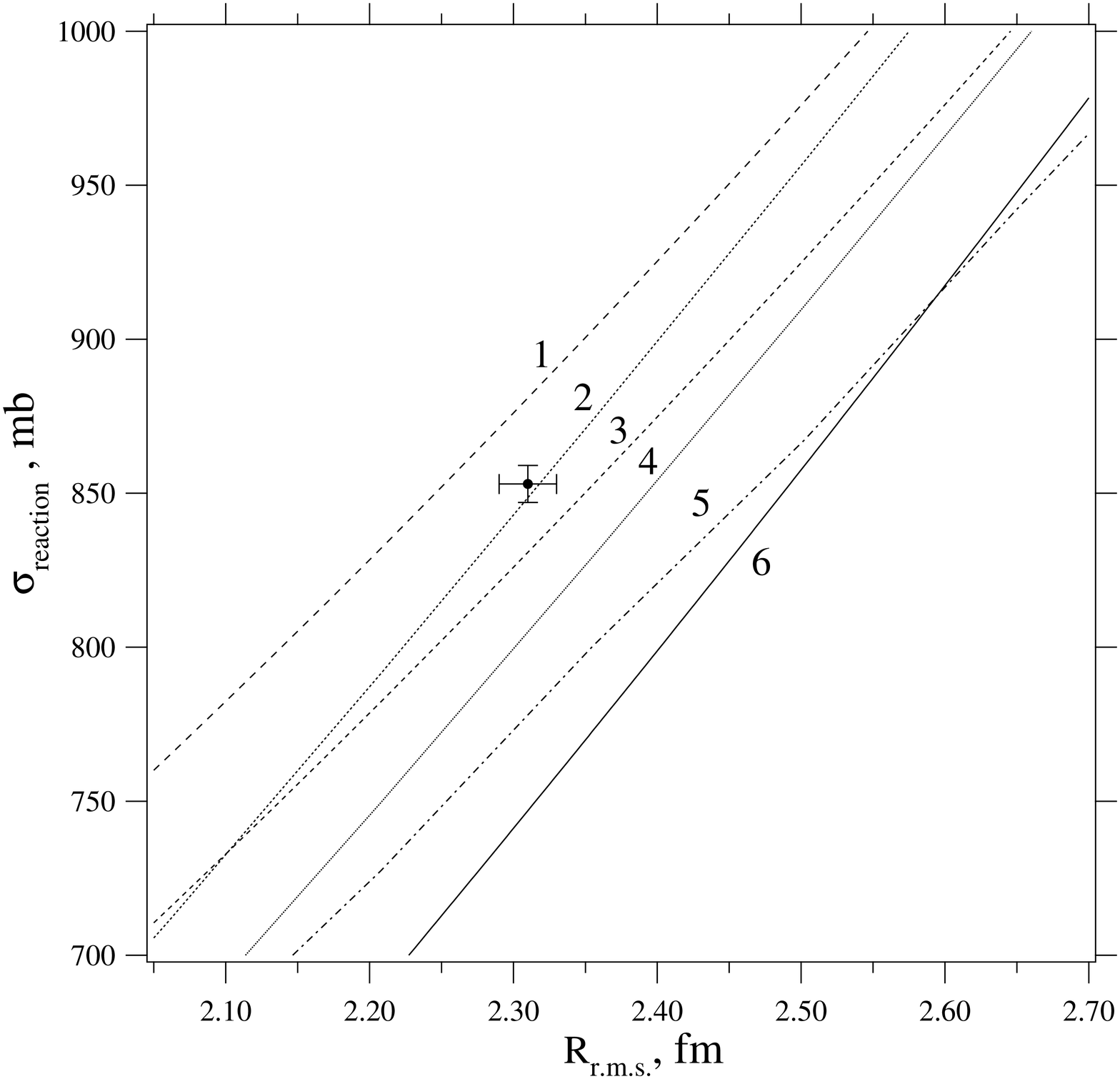,width=0.7\textwidth}} \\
\end{center}
\vskip-0.1cm
Fig.~3. The cross section of the reaction $^{12}$C$-^{12}$C as a function of the ${\rm R_{rms}}$ of the $^{12}$C nucleus calculated in different approximations with two different nucleon densities: 1) the optical approximation without range of NN interaction with Woods-Saxon density distribution; 2) the optical approximation without range of NN interaction with HO-potential density distribution; 3) the rigid target approximation without range of NN interaction with Woods-Saxon density distribution; 4) the rigid target approximation without range of NN interaction with HO-potential density distribution; 5) Glauber calculation with Woods-Saxon density distribution; 6) Glauber calculation with HO-potential density distribution. Calculation in ref.~\cite{Oza} is shown as a marker (\textbullet) for comparison.
\end{figure}

As the first step we have reproduced the result from ref.~\cite{Oza} for $\sigma^{(r)}_{^{12}C-^{12}C}$ in the optical approximation without NN interaction range, using the harmonic oscillator density. Dependence of the  reaction cross section, $\sigma^{(r)}_{^{12}C-^{12}C}$, on the ${\rm R_{rms}}$ is shown by curve 2 of Fig. 3. It is in a good agreement with the result from ref.~\cite{Oza}, which is shown by marker. 

Curve 1 in Fig. 3 represents the calculations of the dependence of the reaction cross section $\sigma^{(r)}_{^{12}C-^{12}C}$ on ${\rm R_{rms}}$ which was done in the optical approximation with the Woods-Saxon density distribution.  By comparing curves 1 and 2 of Fig.~3 one can see that equal ${\rm R_{rms}}$ with different assumptions about the nuclear density distribution result in different reaction cross sections. In other words this equivalently means that the same experimental reaction (or interaction) cross section with different assumptions about nuclear density distribution result in different ${\rm R_{rms}}$, e.g. the assumption of the Woods-Saxon density distribution in $^{12}$C nucleus leads to a smaller value of ${\rm R_{rms}}$ than the one obtained with harmonic oscillator density distribution.
 
The corresponding results obtained in the rigid target approximation are shown by curves 4 and 3 in Fig.~3. Here again the assumption of Woods-Saxon density distribution results in a smaller value of ${\rm R_{rms}}$ than the one calculated with the HO density distribution. The rigid target approximation contains additional diagrams Figs.~1d, 1e, 1f, 1g,..., which increase the shadow effects. That explains why both curves 3 and 4 lie below curves 1 and 2. 

The curves 5 and 6 in Fig.~3 show the results of the calculation of all the diagrams of the Glauber Theory by using Monte Carlo method and accounting for the finite range of NN interaction. Curves 5 and 6 were calculated with the Woods-Saxon density distribution and the HO density respectively. Here, new shadow corrections of the type shown in Figs.~1h, 1i, 1j, 1k,..., appear by comparison to the rigid target approximation. As a result, the calculated reaction cross section becomes now smaller at the same value of ${\rm R_{rms}}$. 
 
The values of the interaction cross sections $\sigma^{(I)}_{^{12}C-^{12}C}$ presented in ref.~\cite{Oza} are $856 \pm 9$ mb and $853 \pm 6$ mb at energies 790 MeV and 950 MeV per nucleon, respectively. The older experimental measurement in ref.~\cite{Jar} gives a value $\sigma^{(I)}_{^{12}C-^{12}C} = 939 \pm 49$ mb at energy 870 MeV per nucleon, i.e. a little larger cross section. The total $^{12}$C-$^{12}$C cross section was measured to be $1254 \pm 54$ mb at the same energy , whereas the Glauber Theory with ${\rm R_{rms}}$ taken from data in ref.~\cite{Oza} predicts a value $\sigma^{tot}_{^{12}C-^{12}C}$ = 1405 mb.
 
The ${\rm R_{rms}}$ values extracted from the measurement of the interaction cross sections of stable projectile nuclei at energies 800$-$1000 MeV scattered from $^{12}$C target (see refs.~\cite{Oza,Oza1}) are shown in Table 1.  The ${\rm R_{rms}}$ values were calculated assuming that the nuclear matter density distribution can be described by the Woods-Saxon expression Eq.~(18). To get the dependence of the reaction cross section $\sigma^{(r)}$ on the ${\rm R_{rms}}$, we varied parameter $c$ of the density distribution and kept parameter $a$ as a constant at $a = 0.54$ fm. The ${\rm R_{rms}}$ values were extracted from the agreement of the calculated value $\sigma^{(r)}$ with the experimental values of $\sigma^{(I)}$.

In Table~1 one can see that our calculations in the optical approximation with the Woods-Saxon density distribution and neglecting the NN interaction range result in the slightly smaller values of ${\rm R_{rms}}$ (0.05$-$0.1 fm) than those obtained in refs.~\cite{Oza,Oza1}.   ${\rm R_{rms}}$ values are getting even smaller when calculated with a finite range of NN interaction. In the case of the Glauber Theory with the Woods-Saxon density distribution shadow corrections lead to larger values of ${\rm R_{rms}}$ than in the other calculations.

\begin{center}
\vskip 5pt
\begin{tabular}{|c||c|c|c|c|c|} \hline
Nucleus & \multicolumn{2}{c|}{Without NN range} & 
\multicolumn{2}{c|}{With NN range} & Glauber Theory\\ \cline{2-5}
& HO, optical, \cite{Oza,Oza1} & WS, optical & WS, optical & WS, rigid target &  \\   \hline
C$^{12}$ & $2.31 \pm 0.02$ & $2.25 \pm 0.01$ & $2.09 \pm 0.01$ & $2.18 \pm 0.01$ & $2.49 \pm 0.01$ \\ 
N$^{14}$ & $2.47 \pm 0.03$ & $2.42 \pm 0.03$ & $2.23 \pm 0.03$ & $2.35 \pm 0.04$ & $2.64 \pm 0.03$ \\ 
O$^{16}$ & $2.54 \pm 0.02$ & $2.48 \pm 0.02$ & $2.29 \pm 0.02$ & $2.41 \pm 0.03$ & $2.69 \pm 0.02$ \\ 
F$^{19}$ & $2.61 \pm 0.07$  & $2.55 \pm 0.08$ & $2.34 \pm 0.08$ & $2.44 \pm 0.09$ & $2.75 \pm 0.07$ \\ 
Ne$^{20}$ & $2.87 \pm 0.03$ & $2.84 \pm 0.04$ & $2.63 \pm 0.03$ & $2.75 \pm 0.04$ & $2.99 \pm 0.03$ \\ 
Na$^{23}$ & $2.83 \pm 0.03$ & $2.73 \pm 0.04$ & $2.52 \pm 0.04$ & $2.62 \pm 0.04$ & $2.91 \pm 0.03$ \\ 
Mg$^{24}$ & $2.79 \pm 0.15$ & $2.65 \pm 0.23$ & $2.44 \pm 0.22$ & $2.53 \pm 0.24$ & $2.85 \pm 0.20$ \\ 
Cl$^{35}$  & $3.045 \pm 0.037$ & $2.92 \pm 0.04$ & $2.68 \pm 0.04$ & $2.76 \pm 0.04$ & $3.08 \pm 0.04$ \\ 
Ar$^{40}$  & $3.282 \pm 0.036$ & $3.16 \pm 0.04$ & $2.90 \pm 0.03$ & $2.98 \pm 0.04$ & $3.30 \pm 0.03$ \\ 
\hline
\end{tabular}
\end{center}
Table 1. The values of ${\rm R_{rms}}$ extracted from the measurements refs.~\cite{Oza,Oza1} of interaction cross section in collisions of projectile nuclear beam with $^{12}C$ target at energies of 800-1000 MeV per nucleon.

\section{Comparison of ${\rm R_{rms}}$ obtained from nucleus-nucleus collisions to electron scattering data}

Let us compare the values for ${\rm R_{rms}}$ we have obtained from nucleus-nucleus collisions with the published results. It is known from refs.~\cite{CLS,ABV} that radii of proton and neutron distributions in nuclei with $Z \simeq A/2$ are practically equal, so we can compare calculated radii for nuclear matter with electrical charge radii presented in ref.~\cite{DeDe}. 

It is necessary to make a distinction between the distributions of the centres of nucleons $\rho_A(r)$ and the folded distributions $\tilde {\rho}_A(r)$, where density $\rho_A(r)$ is convoluted with the matter or charge density in the nucleon, $\rho_N(r)$:
\begin{equation}
\tilde{\rho}_A(r) = \int  \rho_A(r-r_1) \rho_N(r_1) d^3r_1 \;.
\end{equation}

\noindent
We have to deal with $\rho_A(r)$ and with $\tilde{\rho}_A(r)$ when we calculate ${\rm R_{rms}}$  with and without accounting for the range of NN interaction, respectively. The r.m.s. radii of $\tilde{\rho}_A(r)$ and $\rho_A(r)$, $\tilde{\rm R}_{rms}$ and ${\rm R_{rms}}$, are different and the following relation between them was used in ref.~\cite{CLS}:
\begin{equation}
\tilde{\rm R}_{rms}^2 = {\rm R_{rms}^2} +(0.82 \hspace{0.1cm}{\rm fm})^2 \;.
\end{equation}

In Fig.~4 we compare the result of ${\rm R_{rms}}$ calculations in the optical approximation (triangles) and in the rigid target approximation (squares) with $\tilde{\rm R}_{rms}$ values extracted from the electron-nucleus scattering experiment, ref.~\cite{DeDe}.  Calculations of  ${\rm R_{rms}}$  in both approximations were done with zero range of NN interaction (i.e. the folded distribution $\tilde{\rho}_A(r)$ was used) and  Woods-Saxon density distribution. Obtained results are systematically smaller than the data presented in ref.~\cite{DeDe}. This supports our point of view that in the optical approximation and in the rigid target approximation the effects of nuclear shadowing are too small, and, thus, one obtains agreement to the experimental nucleus-nucleus cross section with a smaller value of $\tilde{\rm R}_{rms}$ .
\begin{figure}[htb]
\begin{center}
\mbox{\psfig{file=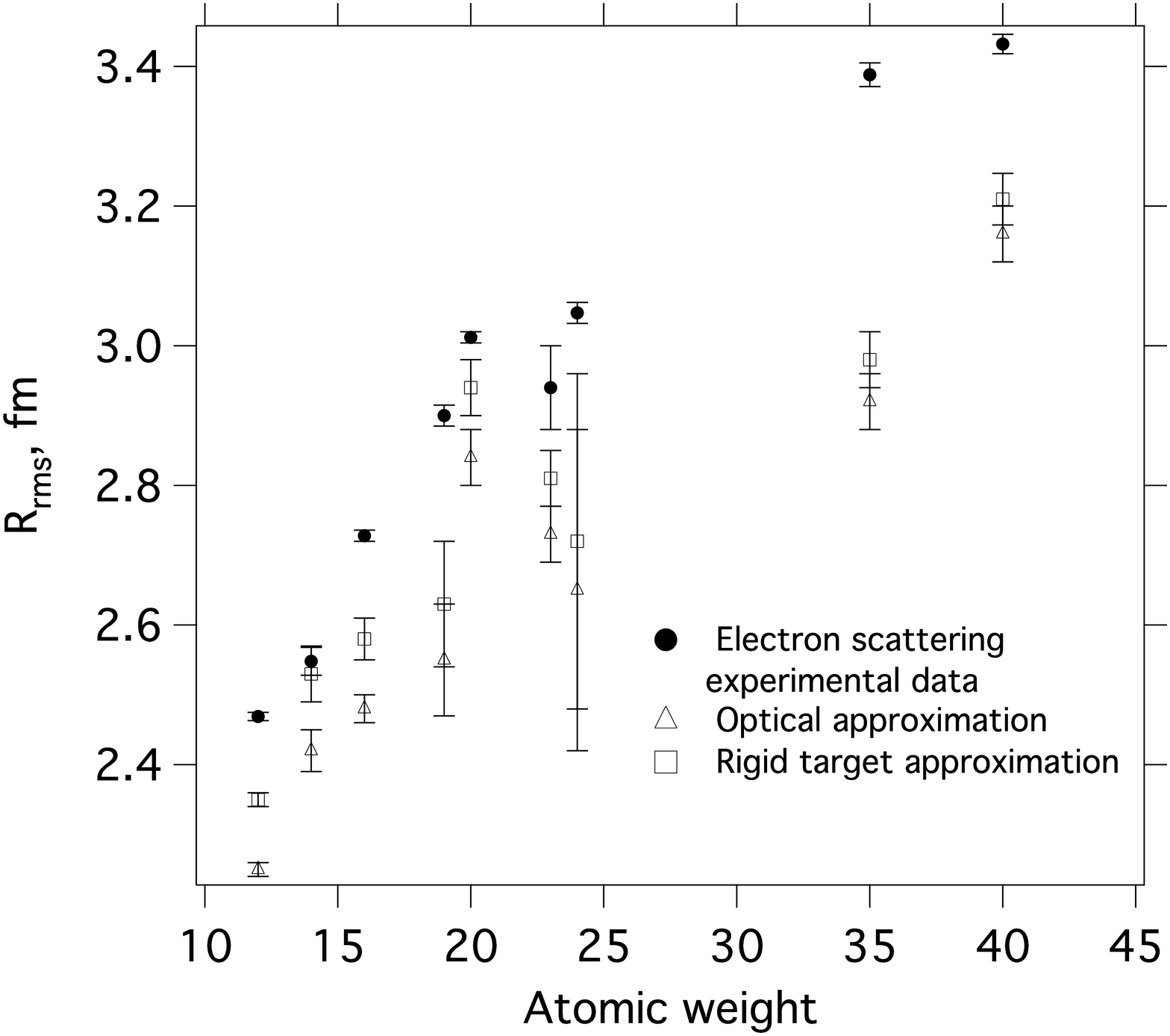,width=0.7\textwidth}} \\
\end{center}
Fig.~4. The values of ${\rm R_{rms}}$ extracted from electron-nucleus scattering  ref.~\cite{DeDe} (filled circles), from nucleus-nucleus collisions in the optical approximation  (triangles), and in the rigid target approximation (squares), both with zero range of NN interaction and with Woods-Saxon density distribution.
\end{figure}

The same calculations of ${\rm R_{rms}}$ were done in the framework of Glauber Theory. The results of these calculations are presented in Fig.~5. In this case it is impossible to provide calculations with zero range of NN interaction because the contributions of diagrams with loops (see for example, Fig.~1k) present this range in the denominator. 

The ${\rm R_{rms}}$ values calculated with distribution $\rho_A(r)$ together with the $\tilde{\rm R}_{rms}$ values extracted from electron-nucleus scattering experiment are presented in the left panel of Fig.~5.  They are in slightly better agreement to electron-nucleus scattering data than in the case shown in Fig.~4. 
\begin{figure}[htb]
\begin{center}
\mbox{\psfig{file=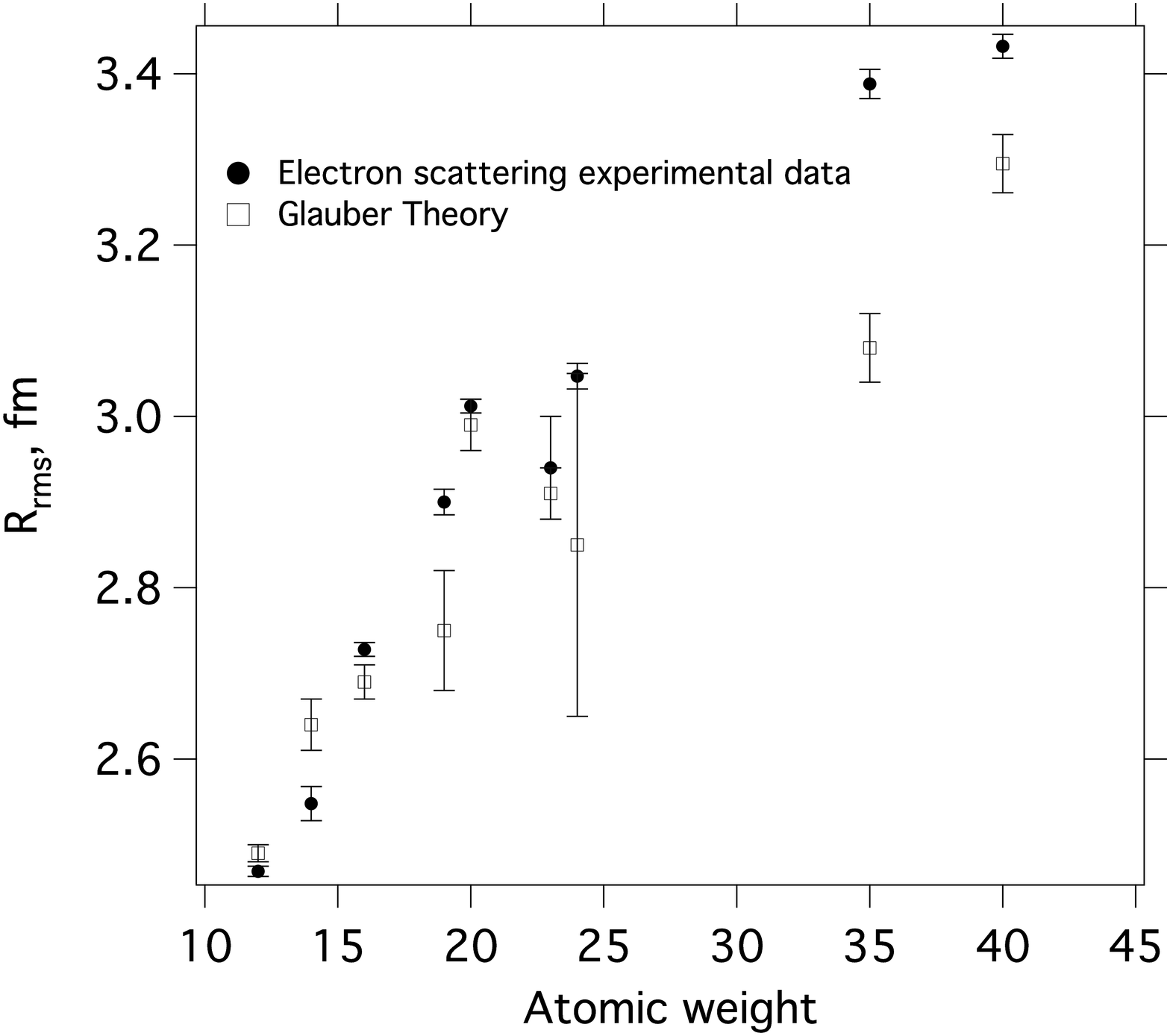,width=0.48\textwidth}}
\mbox{\psfig{file=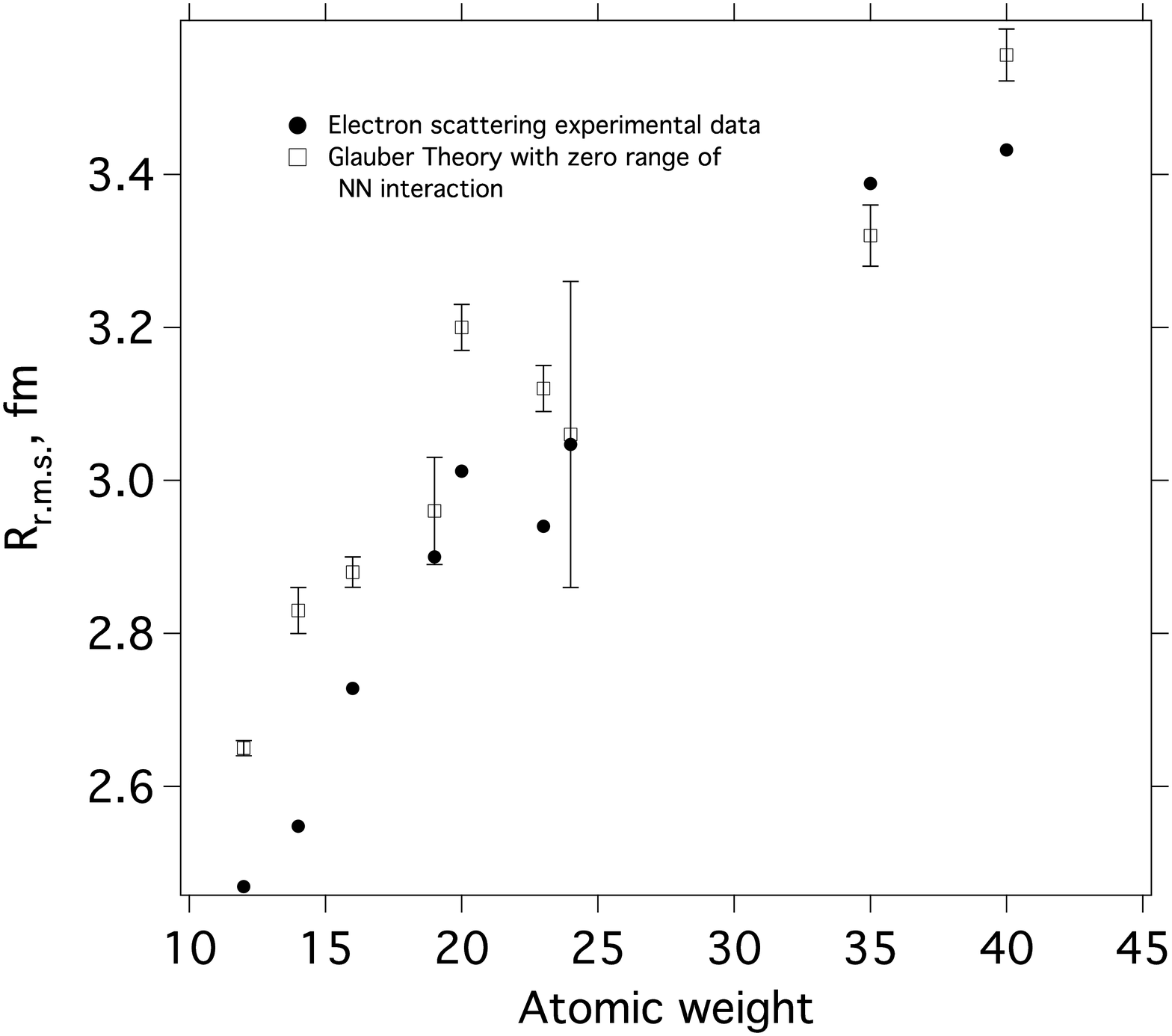,width=0.48\textwidth}} \\
\end{center}
Fig.~5. The values of ${\rm R_{rms}}$ extracted from electron-nucleus scattering  ref.~\cite{DeDe} (filled circles) and from nucleus-nucleus collisions in the Glauber Theory (open square) with Woods-Saxon density distribution and with finite range of NN interaction (left panel) and with zero range of NN interaction (right panel).
\end{figure}

However, as we discussed before,the electron scattering data are related to the folded distributions  $\tilde {\rho}_A(r)$ \footnote{We neglect the difference of electromagnetic and strong interaction nucleon radii.}.

In order to make a more reasonable comparison, we calculate the needed ${\rm \tilde{R}_{rms}}$ in the case of the Glauber Theory as ${\rm R_{rms}} + \Delta$,  where $\Delta$ was calculated as a difference between the ${\rm R_{rms}}$ obtained with the distribution of the nucleon centers (see for example Eq.~18) and ${\rm R_{rms}}$ obtained with the nuclear matter density distribution, Eq.~20.  Both root-mean-square radii were calculated in the optical approximation.

The results are shown in the right panel of Fig.~5. The agreement for not very light nuclei (A $>$ 20) seems to be reasonable. At A $< 20$ the Glauber Theory overestimates the radii, what probably means that the assumption about Woods-Saxon density distribution is not good enough for these nuclei. One can see from Fig.~3 that using of the HO density distribution results in smaller r.m.s. values for a given nucleus-nucleus cross section. 

\section{Conclusion}

We have shown (see Fig.~3) that the calculations of reaction cross sections for nucleus-nucleus collisions with the same ${\rm R_{rms}}$, but different assumptions about nuclear density distribution leads to different results.

These results depend on the used theoretical approximation, the numerical differences among the values of reaction cross sections for nucleus-nucleus collisions obtained with the same nuclear density distribution in the optical approximation, in the rigid target approximation, and in the Glauber Theory, being rather significant, $\sim 0.05 - 0.1$ fm.

The nuclear radii obtained from nucleus-nucleus interactions in the Glauber Theory are in better agreement with the electron-nucleus data than the ones obtained in the optical approximation and in the rigid target approximation. 

We are grateful to G.D. Alkhazov  for useful discussions. This work was supported by Ministerio Educaci\'on y Ciencia of Spain under project FPA 2005--01963 and by Xunta de Galicia (Spain). It was also supported in part by grant RSGSS--1124.2003.2 and by NSF EPSCoR Program (RSF-023-05).


\end{document}